\documentstyle[12pt]{article}
\newcommand{\be}{\begin{equation}}
\newcommand{\en}{\end{equation}}
\newcommand{\bea}{\begin{eqnarray}}
\newcommand{\ena}{\end{eqnarray}}

\newcommand{\hbo}{\hbox to 1 true cm {\hfill } }
\newcommand{\tr}{\hbox{tr}}
\newcommand{\Tr}{\hbox{Tr}}

\def\dslash{\partial\kern-.5em\slash}
\def\kslash{k\kern-.5em\slash}

\begin{document}
\vglue 1truecm

\vbox{
{T95/110 \hfill September 12, 1995 }
}
\vbox{ hep-ph 9509281 \hfill }

\vfil
\centerline{\large \bf Quark confinement in a random background
   Gross-Neveu model}

\vspace{1 true cm}
\centerline{ Kurt Langfeld$^{*}$ }
\vspace{1 true cm}
\centerline{\it Institute for Nuclear Theory}
\centerline{\it University of Washington}
\centerline{\it Seattle, WA 98195, U.S.A.}
\centerline{and}
\centerline{\it Service de Physique Th\'eorique, C.E.A. Saclay, }
\centerline{\it F--91191 Gif--sur--Yvette Cedex, France. }
%
\vskip 1.5cm

\begin{abstract}
\noindent
We study an extended Gross-Neveu model with $N_f$ quark flavors and with
an additional SU(2) (global color) degree of freedom of the quarks.
The four fermion interaction in the color channel is mediated by a
random color matrix with fixed strength. The effective potential for
the quark condensate and the
scalar correlation function is investigated in the large $N_f$-limit.
Quark anti-quark thresholds are absent in the scalar correlation
function implying that the decay of the scalar meson in free quarks is
avoided.
The transition from the low energy to the high energy perturbative regime
is found to be smooth. Perturbation theory provides a good description
of Green's functions at high energies, although there is no
quark liberation.

\end{abstract}

\vfil
\hrule width 5truecm
\vskip .2truecm
\begin{quote}
$^*$ Supported in part by DFG under contracts La--$932/1-1/2$.
\end{quote}
\eject
\section{ Introduction }

One of the most challenging problems nowadays in hadron physics
is to understand why mesons and baryons do not decay into free quarks
(quark confinement), although these are the fundamental constituents
of Quantumchromodynamics (QCD), the right theory of strong interactions.
Lattice simulations verify the quark confinement from a theoretical
point of view~\cite{wi74,eng87}. More and more evidence arose in the recent
past, that the quark confinement is due to a condensation of monopoles
which subsequently expels electric flux from the
vacuum~\cite{ba91}. This provides a linear rising potential between
static quarks implying confinement. A recent success by Seilberg and
Witten~\cite{sei94,sei94b} shows that monopole condensation is
indeed the mechanism of quark confinement in certain super-symmetric
Yang-Mills theories.

Despite the increasing knowledge of the low energy properties of
QCD, hadron physics is still only feasible by effective quark
models~\cite{nambu,ja92}. They successfully describe the mechanism
of the spontaneous breakdown of chiral symmetry and the physics of the
light mesons, but inherently suffer from a non-confinement of the
quarks~\cite{ja92}. Recently, an effective quark model was proposed
which incorporates the quark confinement in a sense that quark
anti-quark thresholds are absent in (mesonic) Green's
functions~\cite{la95}. The main ingredient in this model is
an effective low energy four quark interaction mediated by a random
gluonic background field. The ground state properties of this model
were investigated under extreme conditions, i.e.\ high temperature
and/or density. The results were found to be in agreement with
the phenomenological expectations.

In order to study further the impact of random background fields on
the quark confinement, we here study an extended version of the
Gross-Neveu model~\cite{gro74}. As in QCD, the effective (running)
coupling strength of this model becomes small at high energies
(asymptotic freedom). Here, we will illustrate a mechanism which
serves as a possible answer to the question how perturbation theory
at high energies provides a good description of Green's functions,
although there is no quark liberation.

\section{Ground state properties }
\subsection{ Model description and effective potential }

The model under investigations is described by the Euclidean
generating functional in two dimensions for mesonic Green's
functions, i.e.
\bea
Z[j] &=&
\left\langle \ln \int {\cal D} q \; {\cal D} \bar{q} \; \exp \left\{
- \int d^{2}x \; [ L \; - \; j(x) \, g \bar{q}(x) q(x)  ] \, \right\}
\right\rangle _{O} \; ,
\label{eq:1} \\
L &=& \bar{q}(x) \left( i\dslash + im \right) q(x) \, + \,
\frac{ g }{2 N_f } (\bar{q}(x) q(x) )^{2} \, + \,
\frac{1}{2 N_f} \bar{q} \tau ^{a} q(x) \, \bar{q} \tau ^b q(x) \,
G^{ab} \; ,
\label{eq:2}
\ena
where $m$ is the current quark mass. In the limit $m=0$, the theory
is invariant under global chiral rotations.
The quarks fields $q(x)$ transform under a global $O(N_f)$ flavor
and a global SU(2) (color) symmetry. $g$ and $G^{ab}$ are dimensionless
coupling constants in the color singlet and the color triplet channel
respectively. An average
of all orientations $O$ of the background field $G^{ab}$,
transforming as $G'=O^{T} G O$ with $O$ being a $3 \times 3$
orthogonal matrix, is understood in (\ref{eq:1}) to
restore global $SU(2)$ color symmetry. From~\cite{la95} we expect
that these randomness leads to the confinement of quarks.

In order to explore ground state properties of the model, it is
convenient to study the effective potential $U$ of the (Euclidean)
condensate, since a non-vanishing value signals the spontaneous breakdown
of chiral symmetry.
The effective action $\Gamma ([g\, \bar{q}q]_c)$ of the fermion condensate
is obtained by a Legendre transformation of $Z[j]$
with respect to $j$, i.e.
\be
\Gamma ([g \, \bar{q}q]_c) := - Z[j] + \int d^{2}x \;
 [g \, \bar{q}q]_c (x)  j(x) \; , \hbo
[g \, \bar{q}q]_c (x) :=  \frac{ \delta \, Z[j] }{ \delta j(x) } \; .
\label{eq:e0}
\en
The effective potential $U$ results from $\Gamma $ by restricting
$[g \, \bar{q}q]_c (x)$ to constant values, i.e.
$U([g\bar{q}q]_c )=\Gamma ([g \, \bar{q}q]_c)/V $, where $V$ is the
Euclidean four volume. The potential $U$ is determined up to an unphysical
offset, which is usually chosen in order the potential $U$
to be zero at vanishing condensate $[g \, \bar{q}q]_c$.
The value of the quark condensate at
which the effective potential attaches its global minimum is the
vacuum expectation value $\langle g \, \bar q q \rangle $.
The corresponding minimal value of $U$ is the vacuum energy density.
Serveral minima correspond to different phases of the model.
The state with the lowest energy density (value of $U$ at the local
minimum) forms the true vacuum.

\subsection{ Solution of the model in the large $N_f$-limit }

Although an exact solution of the model might be feasible with the
powerful techniques developed in the context of the
Gross-Neveu model~\cite{be78}, it is sufficient for our purposes
to study the model (\ref{eq:1}) in leading order of a $1/N_f$-expansion.
In the context of two-dimensional models, the $1/N$-expansion was
cast into doubt at least for models with a continuous symmetry like
the chiral Gross-Neveu model, since it predicts a spontaneous
breakdown of the symmetry. The occurrence of Goldstone
bosons in two dimensions causes infra-red problems implying that
the symmetry cannot be broken spontaneously~\cite{col73}.
However, it was subsequently shown by Witten that the symmetry is
``almost'' spontaneously broken, and that the large $N$-expansion provides
a rather good guide to the ground state properties~\cite{wi78}.

The $1/N_f$-expansion is most easily derived by rewriting the
theory (\ref{eq:1}) by means of auxiliary fields , i.e.
\bea
Z[j] &=&
\left\langle \ln \int {\cal D} q \; {\cal D} \bar{q} \;
{\cal D} \Sigma _0 \; {\cal D } \Sigma ^a \; \exp \left\{
- \int d^{2}x \; [ L_M \; - \; j(x) \, g \bar{q}(x) q(x)  ] \, \right\}
\right\rangle _{O} \; ,
\label{eq:3} \\
L_M &=& \bar{q}(x) \left( i\dslash + i\Sigma _0 + i \Sigma ^a \tau ^a
\right) q(x) \, + \,
\frac{ N_f }{2 g } (\Sigma _0 - m )^{2} \, + \,
\frac{N_f}{2} \Sigma ^{a} (G^{-1})^{ab} \Sigma ^b  \; .
\label{eq:4}
\ena
Integrating out the quarks in (\ref{eq:3}) one obtains an equivalent
theory formulated in the fields $\Sigma _0$, $\Sigma ^a$ only.
The non-perturbative approach applied here is an expansion with
respect to $1/N_f$ for a fixed background field $G^{ab}$ and an
average over all orientations $O$ of $G^{ab}$ once the Green's function
has been calculated. For this purpose, we decompose the auxiliary fields
into a mean-field part and fluctuations, i.e.
\be
\Sigma _0 (x) \; = \; M_0 \, + \, \sigma _0 (x) \; , \hbo
\Sigma ^a (x) \; = \; i M^a \, + \, \sigma ^a (x) \; ,
\label{eq:5}
\en
where the orientation of the mean-field $M^a$ depends on the actual
choice of $G^{ab}$. The imaginary unit in front of $M^a$ in (\ref{eq:5})
was introduced for convenience. Later it will turn out that
the ground state favors an imaginary constituent quark mass
(corresponding to real $M^a$) in the color triplet channel.
This will be the main observation implying the absence of quark
anti-quark thresholds in mesonic correlation functions.

We will evaluate the functional integral in (\ref{eq:3}) in a
semi-classical expansion around the mean field values.
(It was recently pointed out that mean-field theory might generically
provide a good description of the ground state of fermionic theories
even at finite temperatures~\cite{ko95}).
As in the case of the standard Gross-Neveu model~\cite{gro74},
fluctuations $\sigma _0$, $\sigma ^a$ are suppressed
by a factor $1/N_f$ implying that we might confine ourselves to the
mean-field level (as long as ground state properties are addressed).
For constant external sources $j$, it is sufficient to consider
constant mean field values. A straightforward calculation yields
\bea
- \frac{1}{VN_f} Z[j] &=& - \frac{1}{V} \Tr \ln \left(
i \dslash + i(M_0 + i M^a \tau^a ) \right)
\label{eq:e1} \\
&+& \frac{ 1 }{2 g } (M_0 - m - ig j)^{2} \, - \,
\frac{ 1 }{2} M^{a} (G^{-1})^{ab} M^b  \; ,
\nonumber
\ena
where the trace extends over internal indices as well as over space
time. The master fields $M_0$ and $M^a$ are solutions of
the gap equations
\be
\frac{ \partial Z[j] }{ \partial M_0 } \; = \; 0 \; , \hbo
\frac{ \partial Z[j] }{ \partial M^a } \; = \; 0 \; .
\label{eq:e2}
\en
It is easy to check that the right hand three equations reduce to
an eigenvector equation to determine $M^a$, i.e.
\be
G^{ab} \, M^{b} \; = \; g_c \, M^{a} \; , \hbo
M:= \sqrt{M^aM^a} \; ,
\label{eq:6}
\en
and an equation to calculate $M$. There are three eigenvalues of the
matrix $G^{ab}$. With $g_c$ we define the eigenvalue of $G^{ab}$
which subsequently provides the phase with the lowest action.
Using (\ref{eq:6}), the trace term in (\ref{eq:e1}) can be
reduced to
\bea
&-& \frac{1}{4 \pi } \int _0^{\Lambda ^2} du \;
\ln \left( u^2 + 2 u (M_0^2-M^2) + (M_0^2+M^2)^2 \right)
\; = \; const.
\label{eq:e3} \\
&& - \frac{1}{4\pi } \left\{ 2 A_- + \sqrt{ A_+^2-A_-^2}
\left[ \pi - 2 \hbox{arctan} \frac{ A_- }{ \sqrt{A_+^2-A_-^2} } \right]
\, - \, A_- \, \ln \frac{A_+^2}{\Lambda ^4 } \right\} \; ,
\nonumber
\ena
where $A_{\pm} = M_0^2 \pm M^2$ and $\Lambda $ is the O(4)-invariant
cutoff to regularize the momentum integration.

In order to renormalize the model, one observes that the logarithmic
divergence in (\ref{eq:e3}) is proportional to $A_-$ and
can be therefore absorbed by a redefinition of the coupling
constants, i.e.
\be
- \frac{1}{\pi } \ln \frac{ \Lambda ^2 }{ \mu ^2 } \, + \,
\frac{1}{g} \; = \; \frac{1}{g_R} \; , \hbo
- \frac{1}{\pi } \ln \frac{ \Lambda ^2 }{ \mu ^2 } \, + \,
\frac{1}{g_c} \; = \; \frac{1}{g_R^c} \; ,
\label{eq:8}
\en
In order to completely renormalize the theory, we also define the
renormalized mass by
\be
\frac{m}{g} \; = \; \frac{m_R}{g_R} \; .
\label{eq:e4}
\en
Field renormalization is not requested at the present stage of
approximation.
The $\beta $-function is in either case of the coupling constants
$\beta (g_R) = - 2/\pi \, g_R^2$ and signals asymptotic freedom.
The renormalized coupling constants $g_R, g_R^c$ decrease by
increasing momentum scale $\mu $. Note that in coincidence the
bare coupling constants vanish in the infinite cutoff limit.
This implies that the terms $m^2/g$, $gj$, $mj$ approach zero for
$\Lambda \rightarrow \infty $.

We are now ready to perform the Legendre transformation (\ref{eq:e0}).
The calculation is straightforward. The final result for the
effective potential of the quark condensate in the large $N_f$-limit is
\bea
\frac{1}{N_f} U &=& \frac{1}{2 g_R} M_0^2 - \frac{ m_R }{g_R}
M_0 - \frac{1}{2 g_R^c} M^2 \, + \, \frac{ M_0^2 -M^2 }{2 \pi }
\left( \ln \frac{ M_0^2 + M^2 }{ \mu ^2 } -1 \right)
\label{eq:e5} \\
&-& \frac{ M_0 M }{2\pi } \left( \pi - 2 \, \hbox{arctan} \,
\frac{ M_0^2-M^2 }{ 2 M_0 M } \right) \; ,
\nonumber
\ena
where $M_0$ is directly related to the quark condensate by
\be
M_0 \; = \; - \frac{i}{N_f} \, [g \, \bar{q}q ]_c \; ,
\label{eq:e6}
\en
and $M$ satisfies the gap-equation
\be
\frac{\pi }{g_R^c(\mu )} M \, + \,
M \, \left\{ \ln \frac{ M_0^2+M^2 }{ \mu ^2 }
\, + \, \frac{M_0}{M} \left( \frac{ \pi }{2} - \, \hbox{arctan} \,
\frac{ M_0^2 -M^2 }{2 M_0 M } \right) \right\} \; = \; 0 \; .
\label{eq:e7}
\en
The equations (\ref{eq:e5},\ref{eq:e6},\ref{eq:e7}) are one of our main
results. We have obtained a renormalization group invariant result,
since a change of the renormalization point $\mu $ can be absorbed by a
redefinition of the renormalized coupling constants.

In order to
remove the superficial dependence of the effective potential $U$ on
the subtraction point $\mu $, we introduce two renormalization
group invariant scales $s_0$, $s_c$ by
\be
\frac{1}{g_R(\mu )} \, + \, \frac{1}{\pi } \, \ln \frac{ s_0^2 }{ \mu ^2 }
\; = \; 0 \; , \hbo
\frac{1}{g_R^c(\mu )} \, + \, \frac{1}{\pi } \, \ln \frac{ s_c^2 }{ \mu ^2 }
\; = \; 0 \; .
\label{eq:10}
\en
With the help of (\ref{eq:10}), the effective potential becomes
\be
\frac{1}{N_f} U \; = \;  - \frac{ m_R }{g_R} M_0 \, + \,
\frac{ M_0^2 }{2 \pi } \left( \ln \frac{ M_0^2 + M^2 }{ s_0^2 } -1 \right)
\, + \,
\frac{ M^2 }{2 \pi } \left( \ln \frac{ M_0^2 + M^2 }{ s_c^2 } + 1 \right)
\; \label{eq:e8}
\en
where
\be
M \, \left\{ \ln \frac{ M_0^2+M^2 }{ s_c ^2 }
\, + \, \frac{M_0}{M} \left( \frac{ \pi }{2} - \, \hbox{arctan} \,
\frac{ M_0^2 -M^2 }{2 M_0 M } \right) \right\} \; = \; 0 \; .
\label{eq:e9}
\en
Thereby equation (\ref{eq:e9}) was used to simplify the expression for
the effective potential.
The two free parameters $g_R$, $g_R^c$ of the original theory are
removed in favor of the two renormalization group invariant
scales $s_0$, $s_c$ (dimensional transmutation). One of these scales,
e.g. $s_0$, can be used as fundamental unit to express dimensionful
quantities. The remaining free parameter is the ratio of these scales,
or $\ln s_0^2/s_c^2$.

Equation (\ref{eq:e9}) is numerically solved for given value of $M_0$.
The potential $U$ is subtracted in order to yield zero at $M_0=0$.
The numerical result for the effective potential $U$ in (\ref{eq:e8})
is shown in figure 1.
Note that $M=0$ is always a solution of (\ref{eq:e9}). In this case,
the effective potential U (\ref{eq:e8}) coincides with that of the
standard Gross-Neveu model (solid line in figure 1).
Note, however, that for sufficiently small
values of $M_0$ a solution with $M\not= 0$ exists. This solution
has lower effective potential (dashed line in figure 1).
The global minimum of this branch therefore constitutes the vacuum.
The effective potential $U$ is smooth at the transition from the phase
with $M\not=0$ to the phase $M=0$, although $M$ varies rapidly
(dot-dashed line in figure 1).

Ground state properties are described by the values of the quark
condensate at which the effective potential attaches its global
minimum. A non-zero value
of $M_0$ corresponds to a non-vanishing quark condensate (see
(\ref{eq:e6})), whereas $M$ controls the
absence of quark anti-quark thresholds in Green's functions as we will
see below.
In order the effective potential to have a minimum, $M_0$ and $M$
must satisfy
\be
\pi \frac{m_R}{g_R} \; = \; M_0 \, \left\{ \ln \frac{ M_0^2+M^2 }{ s_0^2 }
\, - \, \frac{M}{M_0} \left( \frac{ \pi }{2} - \, \hbox{arctan} \,
\frac{ M_0^2 -M^2 }{2 M_0 M } \right) \right\} \; ,
\label{eq:11}
\en
in addition to (\ref{eq:e9}).
Figure 2 shows the lowest action solution. If the overall scale
is chosen to be $s_0$, the only parameter is $\ln s_0^2/s_c^2 $.
One observes that if $s_c$ exceeds a certain value, i.e.
\be
s_c \; > \; 2.71828 \ldots \; s_0 \; ,
\label{eq:13}
\en
a phase with a non-vanishing value of $M$ occurs. For values of $s_c$ less
than this critical value, $M$ is zero and the phase coincides with that
of the standard Gross-Neveu model. In section \ref{sec:scf}, we will discuss
the remarkable properties of this phase.

\subsection{ Quark propagator and analytic structure }

Here, we will discuss the Euclidean quark propagator in the phase
characterized by $M\not= 0$. In momentum space, it is
\be
S(k) \; = \; \frac{1}{ \kslash \; + \; i (M_0 \, + \, i M^a \tau ^a)} \; .
\label{eq:e12}
\en
Introducing the eigenvectors $\vert \pm \rangle $ of the matrix
$M^a \tau^a$, i.e.
\be
M^a \tau^a \, \vert \pm \rangle \; = \; \pm \, M \, \vert \pm \rangle \; ,
\label{eq:e13}
\en
the quark propagator (\ref{eq:e12}) decomposes into two parts with
conjugate complex masses, i.e.
\be
S(k) \; = \; \vert + \rangle \, \frac{1}{ \kslash \; + \; i (M_0 \, + \,
i M)} \, \langle + \vert \; + \;
\vert - \rangle \, \frac{1}{ \kslash \; + \; i (M_0 \, - \,
i M)} \, \langle - \vert \; .
\label{eq:e13a}
\en
The hermiticity and unitarity properties of a theory with a pair of
particles with conjugate complex masses were extensively studied
in several models~\cite{lee69,jan93,ma95}. One finds evidence that
such a theory does not necessarily violate unitarity.

In order to evade the triviality problem~\cite{aiz81} of the scalar
$\phi ^4$-theory, a Higgs model with two scalar degrees of freedom
with complex conjugate masses was proposed~\cite{jan93}.
A consistent field theory is obtained by quantizing the scalar modes
in a Hilbert space with indefinite metric. The model can be described
by an Euclidean path integral, although the Minkowskian functional
integral might not exist due to the presence of the so-called
ghost particles~\cite{jan93}. It is argued that microscopic
acausality effects remain undetectable using realistic wave packets,
and that the S-matrix of the model is unitary~\cite{jan93}.

Recently, the connection of a confining potential and the occurrence
of complex singularities in the propagator of the fundamental
degrees of freedom was investigated in QED3~\cite{ma95}.

{}From the literature, we conclude that a model containing particles
with conjugate complex masses does not necessarily violate
the phenomenological requirements for causality and unitarity of the
S-matrix. In contrast, such a model might be a candidate for
a low-energy theory of QCD describing the confinement of quarks.

In this paper, we will not further question the analytic properties
of the extended version of the Gross-Neveu model studied here.
Although this model lacks an immediate application to hadron physics,
we hope it is simple enough to allow for a clarification of the
phenomenology of particles with conjugate complex masses. These
investigations are left to future work.

\section{ The scalar correlation function }
\label{sec:scf}

In order to focus on the physical impact of the imaginary
constituent quark mass in the color triplet channel, we here
concentrate on the scalar correlation function, i.e.
\be
\Delta (p^2) \; = \;
\int d^{4}x \; e^{-ipx} \; \left\langle g \bar{q}q(x) \;
g \bar{q}q (0) \right\rangle \; = \; \int d^{4}x \; e^{-ipx} \;
\frac{ \delta ^2 \, \ln Z[\phi ] }{ \delta j(x)
\delta j(0) } \vert _{j =0 } \; .
\label{eq:14}
\en
Note that $M_0$ and therefore $g \bar{q}q$ is the renormalization group
invariant and the physical quantity.
Unphysical quark anti-quark thresholds manifest themselves as
imaginary part of the correlation function. In the following, we will
establish the absence of an imaginary part of $\Delta (p^2)$ in the
phase with $M \not= 0$.

In order to obtain $\Delta (p^2)$ in leading order of the $1/N_f$-expansion,
we expand the Lagrangian (\ref{eq:4}) up to second order  in the
fluctuations in (\ref{eq:5}), i.e.
\bea
Z[j] &\approx &
\left\langle \ln \int {\cal D} \sigma _0 \; {\cal D} \sigma ^a \;
\exp \{ - S^{(2)} \} \; \right\rangle _{O} \; ,
\label{eq:e14} \\
S^{(2)} &=& \int \frac{ d^2p }{ (2\pi )^2 } \; \left\{
\frac{1}{2} \sigma (p) \Pi _s^0 (p^2) \sigma (-p) +
\frac{1}{2} \sigma ^\alpha (p) \Pi _s^{\alpha \beta } (p^2)
\sigma ^\beta (-p)
\right. \label{eq:e15} \\
&+& \left. i \sigma ^\alpha (p) K^\alpha (p^2) \sigma (-p) -
\frac{ g N_f }{ 2  } \hat{j} (p) \hat{j}(-p)
- i N_f \hat{j} (p) \sigma (-p) \, \right\} \; ,
\nonumber
\ena
where $\hat{j}(p)$ is the Fourier transform of the external source $j(x)$
and
\bea
\Pi _s^0 (p^2) &=& \frac{N_f}{g} \; - \; \int \frac{ d^2k }{ (2\pi )^2 }
\; \tr \left\{ S(k+p) S(k) \right\} \; ,
\label{eq:e16} \\
\Pi _s^{\alpha \beta } (p^2) &=& N_f \, (G^{-1})^{\alpha \beta }
\; - \; \int \frac{ d^2k }{ (2\pi )^2 }
\; \tr \left\{ \tau ^{\alpha } S(k+p) \tau ^{\beta } S(k) \right\} \; ,
\label{eq:e17} \\
K^{\alpha } (p^2) &=&  -i \; \int \frac{ d^2k }{ (2\pi )^2 }
\; \tr \left\{ \tau ^{\alpha } S(k+p) S(k) \right\} \; .
\label{eq:e18}
\ena
The quark propagator $S(k)$, in momentum space, is given by (\ref{eq:e12}).
The calculation of the functions (\ref{eq:e16}-\ref{eq:e18}) is
straightforward and closely follows~\cite{la95}. It turns out
that the quantities (\ref{eq:e16}) and (\ref{eq:e18}) can be expressed in
terms of two functions $H_0(p^2)$, $H_v(p^2)$, i.e.
\be
\Pi_s^0 (p^2) = \frac{N_f}{\pi} \, H_0(p^2) \; , \hbo
K^{\alpha }(p^2) = \frac{N_f}{\pi } \frac{ M^{\alpha } }{ M } \,
H_{v}(p^2) \; .
\label{eq:e19}
\en
One finds
\bea
H_0(p^2) &=& \frac{\pi }{g} \, + \, \frac{1}{2} \int _0^1 d\alpha \;
\ln \frac{ w^2 + 4 M^2 M_0^2 }{ \Lambda ^4 } \, + \, 2 \; ,
\label{eq:15} \\
H_v(p^2) &=& - \int _0^1 d\alpha \; \hbox{arccos} \,
\frac{w}{ \sqrt{w^2 + 4 M^2 M_0^2 } } \; ,
\nonumber
\ena
where
\be
w \; = \; \alpha (1- \alpha ) p^2 \, + \, M_0^2 \, - \, M^2 \; .
\label{eq:e20}
\en
Whereas $H_v$ is finite, $H_0$ needs renormalization. Equation (\ref{eq:8}),
which was employed to renormalize the effective potential, absorbs
the cutoff dependence in (\ref{eq:15}) as expected.
Removing the superficial dependence of $H_0(p^2)$ on the subtraction
point $\mu $ with help of the renormalization group invariant scale
$s_0$ in (\ref{eq:10}), one finally obtains the finite and
explicitly renormalization group invariant result
\be
H_0(p^2) \; = \; \frac{1}{2} \int _0^1 d\alpha \;
\ln \frac{ w^2 + 4 M^2 M_0^2 }{ s_0 ^4 } \, + \, 2 \; ,
\label{eq:e21}
\en
We also find that $M^\alpha $ is an
eigenvector of the polarization matrix $\Pi _s^{\alpha \beta }$, i.e.
\be
\Pi _s^{\alpha \beta } \frac{ M^{\beta } }{M} \; = \; \frac{N_f}{\pi }
\left( \ln \frac{s_0^2}{s_c^2} + H_0(p^2) \right) \;
\frac{ M^{\alpha } }{M} \; ,
\label{eq:e22}
\en
Equipped with the results (\ref{eq:e19}) and (\ref{eq:e22}), it is now
easy to integrate out the meson fields $\sigma _0 $ and $\sigma ^a$
in (\ref{eq:e14}) and to calculate (\ref{eq:14}). The result of this
procedure only depends on $M^2$ (besides $p^2$ and $M_0^2$). This implies
that average over all orientations of the background field
$G^{ab}$ in (\ref{eq:e14}) is trivial, since $M$ is a singlet.
The final result for the scalar correlation function $\Delta (p^2)$ in
leading order of the $1/N_f$-expansion is therefore
\be
\Delta (p^2) \; = \;
- \frac{ \pi  \, N_f }{ H_0 (p^2) \, + \,
\frac{ H_v^2(p^2) }{ H_0(p^2) + \ln s_0^2/s_c^2 } } \; .
\label{eq:e23}
\en
We are now going to study the occurrence of an imaginary part
of the scalar correlation function signaling a quark anti-quark
threshold. For this purpose, we first perform the analytic continuation
of the Euclidean correlation function to Minkowski space. This
continuation follows the lines of~\cite{lee69}.
The correlation function in Mikowski-space $i \Delta _M ( p_M^2) $ as a
function of the Minkowski four-momentum $p_M$ is related to the Euclidean
counterpart by
\be
i \Delta _M (p_M^2) \; = \; \Delta (- p_M^2) \; .
\label{eq:mink}
\en
It is therefore sufficient to study the Euclidean correlation functions
at negative momentum squared.

In order to illustrate the disappearance of the quark anti-quark threshold
for $M \not= 0$, we first study its occurrence for $M=0$. In this case,
our model describes the scalar correlation function in the usual
Gross-Neveu model~\cite{gro74} with a constituent quark mass $M_0$.
The term of interest is the integrand of $H_0(p^2)$ in (\ref{eq:e21}),
which essentially becomes $\ln w/s_0^2$.
This implies that whenever $w$ becomes negative, the function $H_0$,
(\ref{eq:e21}), acquires an imaginary part, which subsequently
describes the decay of the scalar meson in a free quark anti-quark pair.
In order for $w$ to become negative,
the Euclidean momentum $p^2$ must satisfy,
\be
- p^2 \; < \; 4 M_0^2 \; ,
\label{eq:16}
\en
implying that the quark-anti-quark threshold occurs at a
(Minkowskian) momentum
$p_M = 2 M_0$, which is the familiar result. For $M=0$ the function
$H_v(p^2)$ does no harm, since it is identically zero.

In contrast, for $M \not=0$ the functions $H_{0/v}(p^2)$ are real for
all momentum $p^2$. No quark anti-quark threshold occurs at all.
This is our main observation. Figure 3
shows the scalar correlation function as function of $p^2$.
At negative values of $p^2$, a peak occurs near the would-be
threshold position. This peak is enhanced, if the system is driven
towards the deconfinement phase by a choice of the parameter
$\ln s_0^2/s_c^2$. The correlation function smoothly approaches the
perturbative result (dashed line in figure 3)
at large momentum transfer. Our model therefore represents an example
the Green's functions of which can be accurately calculated in
perturbation theory at high momentum, although there is no
quark liberation.

\section{ Conclusions }

An extended version of the Gross-Neveu model~\cite{gro74}
for $N_f$ quark flavors was studied. In addition to the
$O(N_f)$-flavor symmetry, the model possesses a global SU(2)
(say color) symmetry. The interaction in the color triplet channel
possesses a random orientation in color space.
The effective potential for the
quark condensate was calculated in the large $N_f$ limit.
Two phases were found: in one phase, the constituent quark mass
in the color triplet channel is zero $(M=0)$, and a non-vanishing
dynamical generated mass $M_0$ in the color singlet channel occurs.
This phase corresponds to the vacuum phase of the standard
Gross-Neveu model. However, a second phase constitutes the vacuum, since
this phase has lower vacuum energy density. This non-trivial phase
is characterized by an imaginary constituent quark mass in the color
triplet channel ($M\not= 0$). The impact of an imaginary mass
on causality of Green's functions and unitarity of the S-matrix
is discussed in the literature~\cite{lee69,jan93,ma95}.
It is argued that these models do not contradict the phenomenological
requirements. In order to clarify on this issue in the context of
the model presented here, further investigations are needed, which
are left to future work.

Rather than to investigate these theoretical aspects, we here
focus on the remarkable physical consequences of the phase with
$M\not=0$. For this purpose, the scalar correlation function
was studied. In the Gross-Neveu phase $(M=0)$ of the model, an
imaginary part of the correlator corresponding to the quark
anti-quark threshold was found. In contrast, the correlation function
of the phase with $M\not= 0$ is real, and the quark anti-quark
threshold is absent. The occurrence of an imaginary constituent
quark mass in the color triplet channel might therefore serve
as a signal for quark confinement (which is defined as the absence
of quark thresholds in Green's functions).
At large momentum transfer, the scalar correlation function can
be accurately calculated by perturbation theory. The model
provides example that asymptotic freedom is perfectly compatible
with the non-liberation of quarks.

\bigskip
{\bf Acknowledgements: }

I thank professor Mannque Rho for encouragement and support as well
as for helpful discussions.

\newpage
\centerline{ \bf Figure captions }

\vspace{1 true cm}
  Figure 1: The effective potential $U$ and the color triplet constituent
  quark mass $M$ as function of the quark condensate $M_0$ in the chiral
  limit $m_R=0$ for two values of $\ln s_0^2/s_c^2$;
  all quantities are in units of $s_0$.

\bigskip
  Figure 2: The color-singlet and color-triplet constituent quark masses
  $M_0$ and $M$ in the chiral limit $m_R=0$.

\bigskip
  Figure 3: The scalar correlation function as function of the
  Euclidean momentum transfer $p^2$.

\bigskip


\begin{thebibliography}{sch90}
\bibitem{wi74} K.G.\ Wilson, Phys. Rev. {\bf D10} (1974) 2445;
   M. Creutz, Phys. Rev. {\bf D21} (1980) 2308; Phys. Rev. Lett. {\bf 45}
   (1980) 313;  G. M\"unster, Phys. Lett. {\bf B95} (1980) 59;
\bibitem{eng87}{ J.\ Engels, J.\ Jersak, K.\ Kanaya, E.\ Laermann,
   C.\ B.\ Lang, T.\ Neuhaus, H.\ Satz,
   Nucl. Phys. {\bf B280 } (1987) 577.
   A.\ D.\ Giacomo, M.\ Maggiore, S.\ Olejnik,
   Nucl. Phys. {\bf B347 } (1990) 441.
   L.\ D.\ Debbio, A.\ D.\ Giacomo,
   Phys. Lett. {\bf B267 } (1991) 254.
   T.\ L.\ Ivanenko, A.\ V.\ Pochinski, M.\ I.\ Polikarpov,
   Phys. Lett. {\bf B302 } (1993) 458. }
\bibitem{ba91}{ M.\ Baker, J.\ S.\ Ball and F.\ Zachariasen,
   Phys. Rep. {\bf 209} (1991) 73;
   J.\ Hosek, Phys. Lett. {\bf B226} (1989) 377;
   T.\ Banks and M.\ Spiegelglas, Nucl. Phys. {\bf B152} (1979) 478;
   T.\ Suzuki, Prog. Theo. Phys. {\bf 80} (1988) 929; {\bf 81} (1989) 752;
   M.\ Maedan and T.\ Suzuki, Prog. Theo. Phys. {\bf 80} (1988) 929;
   {\bf 81} (1989) 229. }
\bibitem{sei94}{ N.\ Seiberg and E.\ Witten,
   Nucl. Phys. {\bf B426} (1994) 19. }
\bibitem{sei94b}{ N.\ Seiberg and E.\ Witten, Nucl. Phys. {\bf B431}
   (1994)484. }
\bibitem{nambu}{ Y.\ Nambu and G.\ Jona-Lasinio, Phys. Rev. {\bf 124} (1961)
   246,255. }
\bibitem{ja92}{ M.\ Jaminon, R.\ M.\ Galain, G.\ Ripka and P.\ Stassart,
   Nucl. Phys. {\bf A537} (1992) 418;
   R.\ M.\ Galain, G.\ Ripka, M.\ Jaminon and P.\ Stassart,
   Europhys.Lett. {\bf 14} (1991) 7;
   H.\ Weigel, R.\ Alkofer and H.\ Reinhardt,
   Phys. Rev. {\bf D49} (1994) 5958. }
\bibitem{la95}{ K.\ Langfeld and M.\ Rho, {\it Quark Confinement in a
   Constituent Quark Model}, Saclay-preprint T95/070,
   {\bf hep-ph/9506265}. }
\bibitem{gro74}{ D.\ J.\ Gross, A.\ Neveu, Phys. Rev. {\bf D10}
   (1974) 3235. }
\bibitem{be78}{ B.\ Berg, P.\ Weisz, Nucl. Phys. {\bf B146} (1978)205.
   N.\ Andrei, J.\ H.\ Lowenstein, Phys. Rev. Lett. {\bf 43} (1979) 1698.
   C.\ Destri, J.\ H.\ Lowenstein, Nucl. Phys. {\bf B200 } [FS4]
   (1982) 71, {\bf B205} [FS5] (1982) 369.
   P.\ Forgacs, F.\ Niedermayer, P.\ Weisz, Nucl. Phys. {\bf B367}
   (1991) 123-143, 144-157. }
\bibitem{col73}{ S.\ Coleman, Comm. Math. Phys. {\bf 31} (1973) 259. }
\bibitem{wi78}{ E.\ Witten, Nucl. Phys. {\bf B145} (1978) 110. }
\bibitem{ko95}{ A.\ Koci\'c, J.\ Kogut, Phys. Rev. Lett.
   {\bf 74} (1995) 3109. }
\bibitem{lee69}{ T.\ D.\ Lee and G.\ C.\ Wick, Nucl. Phys. {\bf B9}
   (1969) 209. R.\ E.\ Cutkosky, P.\ V.\ Landshoff, D.\ I.\ Olive
   and J.\ C.\ Polkinghorne, Nucl. Phys. {\bf B12} (1969) 281. }
\bibitem{aiz81}{ K.\ G.\ Wilson, Phys. Rev. {\bf B4} (1971) 3184.
   K.\ G.\ Wilson, J.\ Kogut, Phys. Rep. {\bf 12} (1974) 75.
   M.\ Aizenman, Phys. Rev. Lett. {\bf 47} (1981) 886.
   B.\ Freedman, P.\ Smolensky, D.\ Weingarten, Phys. Lett. {\bf B113}
   (1982) 481.
   J.\ Fr\"ohlich, Nucl. Phys. {\bf B200} (1982) 281.
   C. Arag\~ao de Caravalho, C.\ S.\ Caracciolo, J. Fr\"ohlich, Nucl.
   Phys. {\bf B215} (1983) 209. }
\bibitem{jan93}{ K.\ Jansen, J.\ Kuti, C.\ Liu, Phys. Lett. {\bf B309}
   (1993) 119. }
\bibitem{ma95}{ M.\ Maris, {\it Confinement and complex singularities
   in QED3}, DPNU-95-20, {\bf hep-ph/9508323}. }





\end{thebibliography}
\end{document}